\documentclass[10pt, conference, letterpaper]{IEEEtran}
\usepackage[letterpaper, left = 0.68in, right =0.68in, top = 0.7in, bottom = 1in]{geometry}
\def\arxiv{1}
\def\arxivdisclaimer{1} % With 1, the IEEE disclaimer for arXiv is added
\def\isshorter{1}
\def\review{0} % With 1 comments are enabled, with 0 not

\usepackage[noadjust]{cite}
\if\review1
\usepackage{todonotes}
\else
\usepackage[disable]{todonotes}
\fi
\usepackage[acronym]{glossaries}
\usepackage{amsthm}
\usepackage{amsmath, amssymb}
\usepackage{balance}

\usepackage{svg} % added by AF
\usepackage{graphicx} % added by AF
\usepackage{pgfplots} % added by AF
\usepackage{xfrac} % added by AF
\pgfplotsset{compat=newest} % added by AF
\pgfplotsset{
    /pgfplots/layers/Bowpark/.define layer set={
        axis background,axis grid,main,axis ticks,axis lines,axis tick labels,
        axis descriptions,axis foreground
    }{/pgfplots/layers/standard},
    colormap={jet}{
        rgb255(0cm)=(0,0,128);
        rgb255(1cm)=(0,0,255);
        rgb255(3cm)=(0,255,255);
        rgb255(5cm)=(255,255,0);
        rgb255(7cm)=(255,0,0);
        rgb255(8cm)=(128,0,0)
    }
} % added by AF

\if\arxiv0
\else
\usepackage{hyperref} % [hidelinks] option to get rid of ugly boxes

\glsdisablehyper
\fi
\usepackage{cleveref}
\usepackage{subcaption}
\usepackage{makecell} % For \Xhline from Hr He

\usepackage{textcase}
\usepackage[tablename=TABLE,font=small]{caption}
\DeclareCaptionTextFormat{up}{\MakeTextUppercase{#1}}
\newcommand\blfootnote[1]{%
  \begingroup
  \renewcommand\thefootnote{}\footnote{#1}%
  \addtocounter{footnote}{-1}%
  \endgroup
}

\newtheorem*{remark*}{Remark}

\begin{document}
\title{Antenna Array Design for Mono-Static ISAC}

\author{

\IEEEauthorblockN{
        Alexander~Felix\IEEEauthorrefmark{1}\IEEEauthorrefmark{2},
        Silvio Mandelli\IEEEauthorrefmark{1},
        Marcus~Henninger\IEEEauthorrefmark{1}\IEEEauthorrefmark{2}, 
        and Stephan ten Brink\IEEEauthorrefmark{2} % <-this % stops a space
        }
        
	\IEEEauthorblockA{
	\IEEEauthorrefmark{1}Nokia Bell Labs Stuttgart, 70469 Stuttgart, Germany \\
	\IEEEauthorrefmark{2}Institute of Telecommunications, University of Stuttgart, 70569 Stuttgart, Germany \\
	E-mail: alexander.felix@nokia.com}        
\thanks{TBD: Further notes.}}

\maketitle

\newacronym{2D}{2D}{two-dimensional}
\newacronym{5G}{5G}{fifth generation}
\newacronym{6G}{6G}{sixth generation}
\newacronym{awgn}{AWGN}{additive white Gaussian noise}
\newacronym{cfar}{CFAR}{constant false alarm rate}
\newacronym{isac}{ISAC}{Integrated Sensing and Communications}
\newacronym[plural={MRAs}]{mra}{MRA}{minimum redundancy array}
\newacronym{lmf}{LMF}{location management function}
\newacronym{mf}{MF}{management function}
\newacronym{naf}{NAF}{normalized angular frequency}
\newacronym{NRPPa}{NRPPa}{NR Positioning Protocol A}
\newacronym[plural={PSFs}]{psf}{PSF}{point spread function}
\newacronym{rmse}{RMSE}{root mean squared error}
\newacronym{sara}{SARA}{Sampling And Reconstructing Angular domains}
\newacronym{semf}{SeMF}{Sensing Management Function}
\newacronym[plural={TRPs}]{trp}{TRP}{transmission reception point}
\newacronym[plural={ULAs}]{ula}{ULA}{uniform linear array}
\newacronym[plural={URAs}]{ura}{URA}{uniform rectangular array}

\if\isshorter1
\begin{abstract}
Mono-static sensing operations in \gls{isac} perform joint beamforming between transmitter and receiver. 
However, in contrast to pure radar systems, \gls{isac} requires to fulfill communications tasks and to retain the corresponding design constraints for at least one half-duplex antenna array. This shifts the available degrees of freedom to the design of the second half-duplex array, that completes the mono-static sensing setup of the 6G \gls{isac} system.
Consequently, although it is still possible to achieve the gains foreseen by the radar sparse array literature, it is necessary to adapt these considerations to the new \gls{isac} paradigm.

In this work, we propose a model to evaluate the angular capabilities of a mono-static setup, constrained to the shape of the communications array and its topology requirements in wireless networks.
Accordingly, we enhance the joint angular capabilities by utilizing a sparse element topology of the sensing array with the same number of elements.
Our analysis is validated by simulation experiments, confirming the value of our model in providing system designers with a tool to drastically improve the trade-off between angular capabilities for sensing and the cost of the deployed hardware.

\end{abstract}
% \end{IEEEkeywords}
\else

\fi

\if\arxivdisclaimer1
\blfootnote{This work has been submitted to the IEEE for possible publication. Copyright may be transferred without notice, after which this version may no longer be accessible.}
\vspace{-0.50cm}
\else
\vspace{0.2cm}
\fi

\IEEEpeerreviewmaketitle

\if\isshorter1
\glsresetall
\section{Introduction}

From the start, future \gls{6G} networks promise to further improve performance in terms of throughput, latency and reliability, while at the same time reducing their cost, becoming more energy efficient, and expanding their scope by enabling \gls{isac}~\cite{viswanathan2020communications}. %extending their reach beyond their original scope by enabling \gls{isac}.
Even though \gls{6G} has not been standardized yet, first works have already started to comment on \gls{isac} feasibility and fundamental limits~\cite{mandelli2023survey,liu2022survey}. Backing the considerations up with real world results, a first \gls{isac} demonstrator based on mmWave communications hardware has been published in~\cite{wild2023integrated}.

We consider mono-static sensing setups, defined as the joint sensing operations of a co-located transmitter and receiver, which can optionally be two almost co-located half-duplex arrays, as in~\cite{wild2023integrated}. The determination of the beamforming coefficients of each array is crucial to be able to optimize the joint angular capabilities~\cite{hoctor1990unifying}.
To that end, several algorithmic approaches exist~\cite{Yu2016AltMinAlgo, wang2022beamforming_riemannian,rajamaki2020hybrid}.

It has been shown that sparse arrays with element spacing greater than half the wavelength can improve the angular capabilities of the mono-static setup with a constant number of full-duplex elements~\cite{Koc2022HybridBeamFD},~\cite{Hu2023SparseFD}. 
However, wireless communications need to avoid grating lobes upon signal transmission to mitigate intra- and inter-cell interference. This requires having element spacing at around half the wavelength in communications arrays~\cite{mandelli2023survey}. 

Our contribution starts by realizing that the mentioned half-wavelength constraint does not prevent the improved angular capabilities promised by the sparse array literature~\cite{Koc2022HybridBeamFD,Hu2023SparseFD}. This stems from the half-duplex limitation of communications radios, that can only transmit or receive at the same time. Therefore, an \gls{isac} mono-static setup requires a separate half-duplex receive array focused solely on sensing. 
In a nutshell: differently from full-duplex sparse array designs, we focus on the sparse design of a half-duplex receiver, dissimilar from the transmitter.

In our work, we propose how to design a sensing receive array to optimize the mono-static setup's angular capabilities, without modifying the communications array that will act as a transmitter during sensing operations. Assuming a typical communications \gls{ura}, we model how the resulting angular capabilities of the mono-static setup scale as function of the optimized receive array design.
This avoids any impact on communications performance and -- thus -- the corresponding radio design.
Note that the split of the arrays is compatible with existing \gls{isac} proof of concepts based on communications hardware as in~\cite{wild2023integrated}, where we would only substitute the receive array with our new half-duplex sparse elements arrangement.

Our simulation study investigates the improved angular capabilities from joint beamforming of different mono-static setups. 
The results of our study show that mono-static setups with fewer elements and larger spacing can achieve the same angular capabilities as denser setups, as expected.
Thus, the feasibility of sparse sensing arrays without compromising on the angular capabilities is proven. The reduction in the number of elements of the sensing array design allows for the integration of sensing capabilities in communications at a significantly reduced cost, thus, improving the business case of \gls{isac}.

\section{System Model}

We consider a narrowband, mono-static wireless system, like the one depicted in Fig.~\ref{fig:monostat_sketch}.
Even though in practice a wireless system is not narrowband, the effect in terms of beam squinting is usually negligible due to the
bandwidth being small compared to the carrier frequency~\cite{mandelli2023survey}.
There are two options to achieve this mono-static setup: operating a single array in full-duplex mode or co-locating two arrays and operating them in half-duplex mode.
This paper focuses on the latter, where we assume the angular and beamforming perspectives are co-located. This approach has been proven to work in practice with existing communications hardware~\cite{wild2023integrated}. 
Furthermore, the joint beamforming characteristics of the two arrays, as depicted in Fig.~\ref{fig:monostat_sketch}, depend on a virtual structure called the sum co-array~\cite{hoctor1990unifying}, which will be discussed in \ref{sec:sumco}.

Based on existing communications deployments, we consider a \gls{ura} as transmitter with half wavelength element spacing, needed to avoid any grating lobe in beamforming operations.
Additionally, the receiver array is assumed to be a \gls{ura}, as justified later in~\ref{sec:sumco}.

The total number of elements is given by $N=N_x N_z$, where $N_x$ and $N_z$ denote the number of elements in horizontal and vertical array direction, respectively. To ease notation, but without loss of generality, we consider \glspl{ura} of square shape, $N^{(\text{1D})}=N_x=N_z$, and antenna spacing $d$ in both directions. However, the extension of our analysis to the rectangular case is straightforward. Parameters associated with either array type are denoted by the superscript symbol $(\cdot)$. The specific cases of the communications array as the transmitter and the sensing array as the receiver are denoted by superscripts $(\text{c})$ and $(\text{s})$, respectively.
However, they can exchange roles without affecting the validity of our approach.

The matrix $\mathbf{P}^{(\cdot)}=[\mathbf{p}_1^{(\cdot)},\mathbf{p}_{2}^{(\cdot)},...,\mathbf{p}_{N}^{(\cdot)}]^\mathrm{T} \in \mathbb{R}^{N\times 2}$ comprises the normalized element positions, with elements $\mathbf{p}_n^{(\cdot)}~=~\left[\sfrac{x_n}{d}, \sfrac{z_n}{d}\right]^\text{T}$. 
Under far-field assumption, the relative direction vectors are 
\begin{equation}
\mathbf{u}= \left[\cos(\phi)\cos(\theta), {\cos(\phi)}{\sin(\theta)} \right]^{\text{T}} \; ,
\label{eq:DirectionVector}
\end{equation}
where elevation $\phi$ and azimuth $\theta$ denote the incident angles with regard to the arrays~\cite{hoctor1990unifying}.

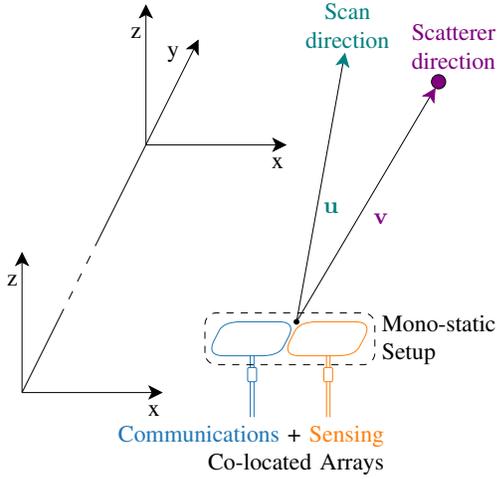
\begin{figure}[t]
    \centering
    \resizebox{0.80\columnwidth}{!}
    {\definecolor{darkorange25512714}{RGB}{255,127,14}
\definecolor{steelblue31119180}{RGB}{31,119,180}
% \definecolor{greenmatplot}{RGB}{44,160,44}
% \definecolor{redmatplot}{RGB}{214,39,40}

\def \globalscale {1.000000}
\begin{tikzpicture}[y=1pt, x=1pt, yscale=\globalscale,xscale=\globalscale, every node/.append style={scale=\globalscale}, inner sep=0pt, outer sep=0pt]
  \path[draw=white,miter limit=10.0] (15.0, 75.8) -- (37.5, 75.8);

  \path[draw=white,miter limit=10.0] (15.0, 83.2) -- (37.5, 83.2);

  \path[draw=white,miter limit=10.0] (15.0, 90.8) -- (37.5, 90.8);

  \path[draw=white,miter limit=10.0] (15.0, 98.2) -- (37.5, 98.2);

  \path[draw=black,miter limit=10.0] (7.5, 38.2) -- (62.7, 38.2);

  \path[draw=black,fill=black,miter limit=10.0] (66.7, 38.2) -- (61.4, 35.6) -- 
  (62.7, 38.2) -- (61.4, 40.9) -- cycle;

  \path[draw=black,miter limit=10.0] (7.5, 38.2) -- (7.5, 93.5);

  \path[draw=black,fill=black,miter limit=10.0] (7.5, 97.4) -- (10.1, 92.2) -- 
  (7.5, 93.5) -- (4.9, 92.2) -- cycle;

  \path (60.0, 38.2) rectangle (67.5, 23.2);

  \begin{scope}[shift={(-0.4, 0.4)}]
    \node[text=black,anchor=south] (text8) at (63.8, 27.8){x};

  \end{scope}
  \path (0.0, 98.2) rectangle (7.5, 83.2);

  \begin{scope}[shift={(-0.4, 0.4)}]
    \node[text=black,anchor=south] (text9) at (3.8, 83.2){z};

  \end{scope}
  \path[draw=black,miter limit=10.0] (7.5, 38.2) -- (25.2, 73.7)(27.3, 
  77.8)(27.3, 77.8) -- (29.0, 81.2)(31.0, 85.3)(31.0, 85.3) -- (32.7, 
  88.7)(34.8, 92.8)(34.8, 92.8) -- (36.5, 96.2)(38.5, 100.3)(38.5, 100.3) -- 
  (80.4, 184.0);

  \path[draw=black,fill=black,miter limit=10.0] (82.1, 187.5) -- (82.1, 181.6) 
  -- (80.4, 184.0) -- (77.4, 184.0) -- cycle;

  \begin{scope}[shift={(-0.4, 0.4)}]
    \node[text=black,anchor=south] (text10) at (72.0, 177.8){y};

  \end{scope}
  \path[draw=black,miter limit=10.0] (60.0, 143.2) -- (115.2, 143.2);

  \path[draw=black,fill=black,miter limit=10.0] (119.2, 143.2) -- (113.9, 140.6)
   -- (115.2, 143.2) -- (113.9, 145.9) -- cycle;

  \path[draw=black,miter limit=10.0] (60.0, 143.2) -- (60.0, 198.5);

  \path[draw=black,fill=black,miter limit=10.0] (60.0, 202.4) -- (62.6, 197.2) 
  -- (60.0, 198.5) -- (57.4, 197.2) -- cycle;

  \path[draw=black,fill=violet] (184, 170) ellipse (3pt and 3pt); %circle scatterer

  \begin{scope}[shift={(-0.4, 0.4)}]
    \node[text=violet,anchor=south west,align=center] (text14) at (173, 175){Scatterer \\direction};

  \end{scope}
  % \path[draw=black,fill=white] (143.2, 170) ellipse (3pt and 3pt); %circle scan

  \path[draw=black,fill=black] (123.8, 68.2) ellipse (1pt and 1pt); %circle panels

  \begin{scope}[shift={(-0.4, 0.4)}]
    \node[text=teal,anchor=south west,align=center] (text15) at (128, 184){Scan \\direction};

  \end{scope}
  \path (112.5, 143.2) rectangle (120.0, 128.2);

  \begin{scope}[shift={(-0.4, 0.4)}]
    \node[text=black,anchor=south] (text16) at (116.2, 132.8){x};

  \end{scope}
  \path (52.5, 203.2) rectangle (60.0, 188.2);

  \begin{scope}[shift={(-0.4, 0.4)}]
    \node[text=black,anchor=south] (text17) at (56.2, 188.2){z};

  \end{scope}
  % \draw[draw=black,dashed,miter limit=10.0] (123.8, 68.2) -- (192.4, 176.7);

  % \path[draw=black,fill=black,miter limit=10.0] (194.5, 180.0) -- (194.0, 
  % 174.2) -- (192.4, 176.7) -- (189.5, 177.0) -- cycle;

  \draw[draw=black,miter limit=10.0] (123.8, 68.2) -- (180.4, 163.7); % arrowline scatterer

  \path[draw=violet,fill=violet,miter limit=10.0] (182.5, 167.0) -- (182.0, 
  161.2) -- (180.4, 163.7) -- (177.5, 164.0) -- cycle; % arrowhead scatterer

  \begin{scope}[shift={(-0.4, 0.4)}]
    \node[text=violet,anchor=south] (text20) at (160.0, 109.5){\(\mathbf{v}\)};

  \end{scope}
  % \path[draw=teal,miter limit=10.0] (123.8, 68.2) -- (141.6, 161.1);

  % \path[draw=teal,fill=teal,miter limit=10.0] (142.3, 164.9) -- (143.9, 
  % 159.3) -- (141.6, 161.1) -- (138.8, 160.3) -- cycle;

  \path[draw=black,miter limit=10.0] (123.8, 68.2) -- (143.5, 177.9); %arrowline scan

  \path[draw=teal,fill=teal,miter limit=10.0] (144.3, 181.9) -- (145.9, 
  176.3) -- (143.6, 178.1) -- (140.8, 177.3) -- cycle; %arrowhead scan

  \begin{scope}[shift={(-0.4, 0.4)}]
    \node[text=teal,anchor=south] (text24) at (138.8, 114.0){\(\mathbf{u}\)};

  \end{scope}
  \path (74.2, 30.8) rectangle (164.2, 0.8);

  \begin{scope}[shift={(-0.4, 0.4)}]
    \node[text=black,anchor=south,align=right] (text25) at (105, 3){\textcolor{steelblue31119180}{Communications} + \textcolor{darkorange25512714}{Sensing}\\Co-located Arrays};
        %     \node[text=black,anchor=south] (text25) at (110, 13){\textcolor{steelblue31119180} {communications} + \textcolor{darkorange25512714}{Sensing}}; % 119.2
    % \node[text=black,anchor=south,align=center] (text25) at (119.2, 3){Co-located Arrays};

  \end{scope}
  \path[draw=steelblue31119180,miter limit=10.0] (105.8, 27.8) -- (105.8, 58.9);

  \path[draw=steelblue31119180,miter limit=10.0] (104.2, 27.8) -- (104.2, 58.9);

  \path[draw=steelblue31119180,fill=white,rounded corners=0.6pt] (103.1, 48.8) rectangle 
  (106.9, 42.8);

  \path[draw=steelblue31119180,fill=white,miter limit=10.0] (100.9, 53.8) -- (93.0, 53.8) 
  .. controls (88.0, 53.8) and (86.7, 56.1) .. (89.0, 60.5) -- (91.0, 64.4) .. 
  controls (92.3, 67.0) and (95.5, 68.2) .. (100.5, 68.2) -- (116.3, 68.2) .. 
  controls (121.3, 68.2) and (122.6, 66.0) .. (120.3, 61.6) -- (118.3, 57.7) .. 
  controls (117.0, 55.1) and (113.8, 53.8) .. (108.8, 53.8) -- cycle;

  \path[draw=darkorange25512714,miter limit=10.0] (138.0, 27.8) -- (138.0, 58.9);

  \path[draw=darkorange25512714,miter limit=10.0] (136.5, 27.8) -- (136.5, 58.9);

  \path[draw=darkorange25512714,fill=white,rounded corners=0.6pt] (135.4, 48.8) rectangle 
  (139.1, 42.8);

  \path[draw=darkorange25512714,fill=white,miter limit=10.0] (133.1, 53.8) -- (125.2, 53.8)
   .. controls (120.2, 53.8) and (118.9, 56.1) .. (121.2, 60.5) -- (123.2, 64.4)
   .. controls (124.6, 67.0) and (127.8, 68.2) .. (132.8, 68.2) -- (148.5, 68.2)
   .. controls (153.5, 68.2) and (154.9, 66.0) .. (152.6, 61.6) -- (150.5, 57.7)
   .. controls (149.2, 55.1) and (146.0, 53.8) .. (141.0, 53.8) -- cycle;

  \draw[draw=black, rounded corners=5pt, dashed] (85.0, 50.0) rectangle ++(72.0, 22.0);
  % \node[text=black,anchor=south west,align=center] (textsumco) at (160.0, 57.0){Sum Co-Array};
  \node[text=black,anchor=south west,align=left] (textsumco) at (160.0, 50.0){Mono-static \\Setup};

\end{tikzpicture}
    }
    \caption{Mono-static setup of co-located communications (Tx) and sensing (Rx) arrays with joint beamforming in a desired scan direction~\(\mathbf{u}\) with a single scatterer at direction~\(\mathbf{v}\).}
    % \todo[inline]{MH: could it make sense to sketch the scatterer somehow, e.g., with a circle or so? Just to make sure that there is something in the environment that reflects; AF: done}
    \label{fig:monostat_sketch}
    \vspace{-3mm}
\end{figure}

\subsection{Joint Beamforming and \glsentrylong{psf}}

For joint beamforming one must define the beamforming coefficient of each individual element $w^{(\cdot)}$.
These coefficients are then aggregated per array in 
\(\mathbf{W}^{(\cdot)}\mspace{-5mu}\in\mspace{-5mu}\mathbb{C}^{N^{(\cdot,\text{1D})} \times N^{(\cdot,\text{1D})}}\). Assuming %narrowband conditions for~
$K$ arbitrary scanned directions, the steering matrix of the array~\(\mathbf{A}^{(\cdot)}\in \mathbb{C}^{N^{(\cdot)}\times K}\)
can be expressed as \(\mathbf{A}^{(\cdot)}\mspace{-9mu}~=~\mspace{-11mu}\left[\mathbf{a}^{(\cdot)}\mspace{-5mu}\left(\mathbf{v}_1\right),...,\mathbf{a}^{(\cdot)}\mspace{-5mu}\left(\mathbf{v}_K\right)\right]\mspace{-13mu}~=~\mspace{-13mu}\left[e^{-jk_0\mathbf{P}^{(\cdot)}\mathbf{v}_1},...,e^{-jk_0\mathbf{P}^{(\cdot)}\mathbf{v}_K}\right]\), with \(k_0 = 2\pi/\lambda\) as the wavenumber and $\lambda$ as the wavelength of the center frequency~\cite{rajamaki2020hybrid}.

The \gls{psf} expresses the spatial impulse response to a single point scatterer in direction \(\mathbf{v}\) of a linear imaging system when the system is steered in a desired scan direction \(\mathbf{u}\):
\begin{equation}
    \psi^{(\cdot)}\mspace{-5mu}\left(\mathbf{u},\mathbf{v}\right)=\mathbf{a}^{(\cdot)}\mspace{-5mu}\left(\mathbf{v}\right)^{\mathrm{T}}\mathrm{vec}\left(\mathbf{W}^{(\cdot)}\mspace{-5mu}\left(\mathbf{u}\right)\right) \; ,
\end{equation}
where $\mathrm{vec}(\cdot)$ denotes the vectorization operator. 
From the \gls{psf} the system's achievable angular capabilities as the resolution, i.e., the width of the main lobe, and the sidelobe suppression can be determined.
For the joint system of the communications and the sensing array, the effective steering effects of all elements need to be considered. This can be derived from the Khatri-Rao product of the steering matrices of both arrays.

\subsection{Antenna Array Model and Sum Co-Array \label{sec:sumco}}

The coherent beamformed signal of a mono-static sensing setup, evaluated for one scan direction vector $\mathbf{u}$, is

\begin{align}
\medmuskip=0mu
\thinmuskip=0mu
\thickmuskip=0mu
\hat{s}\left(\mathbf{u}\right)&=\int\mkern-6mu{s\left(\mathbf{v}\right)\ \psi^{(\text{c})}\left(\mathbf{u}, \mathbf{v}\right)\ \psi^{(\text{s})}\left(\mathbf{u}, \mathbf{v}\right)\ d\mathbf{v}\mathrm{\ } }
\nonumber
\\
\label{eq:monostatic_imaging}
&=
\int\mkern-6mu{s\left(\mathbf{v}\right)\mspace{-2mu} \sum_{n=1}^{N^{(\text{c})}}\mspace{-2mu}\sum_{i=1}^{N^{(\text{s})}}{w_{n}^{(\text{c})}w_{i}^{(\text{s})}e^{-jk_0\left(\mathbf{u}-\mathbf{v}\right)^\text{c}\left(\mathbf{p}_{n}^{(\text{c})}+\mathbf{p}_{i}^{(\text{s})}\right)}}d\mathbf{v}} \; ,
\end{align}
given the angular response of the scenario $s\left(\mathbf{v}\right)$. 
It is evident that the combinations of all element positions $(\mathbf{p}^{(\text{c})} + \mathbf{p}^{(\text{s})})$ determine the angular response of our setup. From these combinations the virtual sum co-array structure is derived~\cite{hoctor1990unifying}, which is denoted with superscript $(+)$ henceforth. 

We recall the Fourier relationship between the element positions and the angular domain~\cite{mandelli2022sampling}. Therefore, in this work we enforce an uniform spacing in the sum co-array element positions to simplify the angular steering operations, that in practice are computed with constant phase shifts across different elements. 
Combined with our focus on communications setups with transmitter \glspl{ura} of $\lambda/2$ spacing, this limits the variations of the receiver sensing array design. Thus, to obtain the $\lambda/2$ spacing in the sum co-array the receiver needs to be an \gls{ura} as well with an element spacing of $d^\text{(s)}=md^\text{(c)}\leq d^\text{(c)}N^\text{(c,\text{1D})}$ with $m\in\mathbb{Z}^+$. We postpone to future work the investigations of different sensing array shapes and the corresponding trade-offs. Under these assumptions, the spacing of the sum co-array elements can be derived as $d^\text{(+)}=\min{\left(d^\text{(c)}, d^\text{(s)}\right)}$ and the number of elements follows as

\begin{align}
N^\text{(+,\text{1D})} = \frac{d^\text{(c)}\left(N^\text{(c,\text{1D})}-1\right)+d^\text{(s)}\left(N^\text{(s,\text{1D})}-1\right)}{d^\text{(+)}} \; ,
\label{eq:sumco_num_elems}
\end{align}
for the virtual structure.

\subsection{Beamforming Considerations from Prior Art}

As demonstrated in \cite{rajamaki2020hybrid}, the achievable \gls{psf} for joint beamforming is determined by the shape of the sum co-array, meaning that different physical arrays can achieve the same angular capabilities. 
As an example, Fig.~\ref{fig:sumco_asy} shows two different setups, where the communications array (Tx) is fixed with \(N^{(\text{c,\text{1D}})}=11\) and \(d^{(\text{c})}=\lambda/2\). The first setup (A) has a sensing array (Rx) of \(N^{(\text{s,\text{1D}})}=9\) and \(d^{(\text{s})}=\lambda/2\), while the second setup (B) has only \(N^{(\text{s,\text{1D}})}=3\) and \(d^{(\text{s})}=2\lambda\). However, in both cases the sum co-array, obtained from the correlation (\(\star\)) of the Tx and Rx element positions, consists of the same \gls{ura} with \(N^{(\text{+,\text{1D}})}=19\) elements and $d^{(+)} = \lambda/2$.

To determine the beamforming coefficients, we adopt the alterating minimization approach of~\cite{rajamaki2019analog, rajamaki2020hybrid}. First, we calculate the desired \gls{psf} using an arbitrary set of beamforming coefficients for the sum co-array. Then, we attempt to achieve the same \gls{psf} by mapping these weights to $Q$ joint transmitter and receiver acquisitions, i.e., \textit{componenent images}, that are summed up. 
The minimum $Q$ required for algorithmic convergence is correlated with the total number of elements~$N$ and the sparseness of the arrays.

\subsection{Normalized Angular Frequency}
When evaluating the \gls{2D} angular response of point scatterers in the radian domain, their observed shape depends on the relative angles to their locations. %Thus, the comparability between different scenarios is reduced.
To ensure invariant point scatterer observations regardless of their position in the sampling space, we utilize the \gls{naf} domain~\cite{mandelli2022sampling} instead of the radian domain.
Therefore, we express the incident angles $\phi$ and $\theta$ as their \gls{naf} equivalents in vertical and horizontal array directions, denoted by $\eta$ and $\ell$, respectively. The equations for the conversions are
\begin{align}
\eta&=\frac{d}{\lambda}\sin \left( \phi \right)
\label{eq:NAF_vertical} \; , \\
\ell&=\frac{d}{\lambda}\sin \left( \theta \right)\cos \left( \phi \right) \; ,
\label{eq:NAF_horizontal}
\end{align}
where $\lambda$ denotes the wavelength of the center frequency.
We sample the \gls{naf} domain in the interval $[-d/\lambda,d/\lambda]$ to avoid aliases, as suggested in~\cite{mandelli2022sampling}.
Note that to cover while beamforming the full angular domain aperture spanned by the (virtual) antenna planar array without aliases, i.e., $\pi$, one should use the well-known formula for the maximum antenna distance, i.e., $d \leq \lambda/2$.

\section{Imaging Setup for \gls{isac} deployments}

In this section, we first introduce prior art beamforming considerations to jointly operate transmit and receive arrays to achieve a desired \gls{psf}. Then, we introduce our proposal to utilize dissimilar arrays in \gls{isac}.

\subsection{Dissimilar Arrays for \gls{isac} Operations}

\begin{figure}
    \centering
    % \textbf{Sum co-array of dissimilar}\par\medskip
    % \includegraphics[scale=0.3]{example-image-a}
    \resizebox{0.9\columnwidth}{!}{
    \input{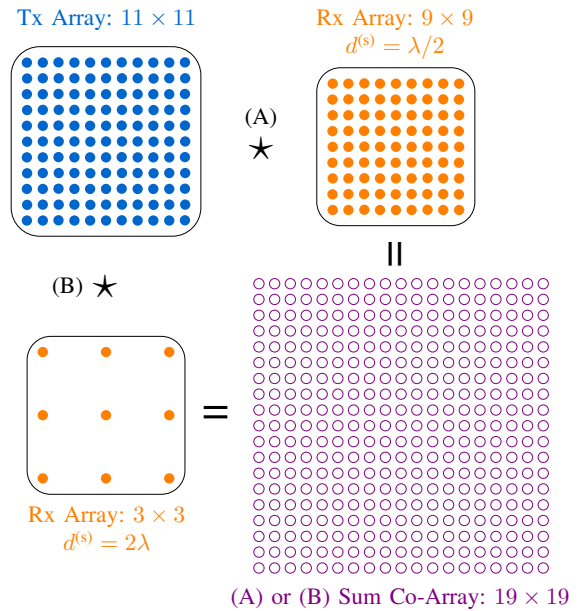}}
    % \caption{Sum co-array of dissimilar arrays derived from communications array with \(\mathcal{S}^{(\text{c})}=11\times 11\) with \(d^{(\text{c})}=\lambda/2\) and sparse sensing array \(\mathcal{S}^{(\text{s})}=3 \times 3\) with \(d^{(\text{s})}=2\lambda\).}
    \caption{Sum co-array of dissimilar \glspl{ura} derived from the correlation~(\(\star\)) of a communications array of \(N^{(\text{c},\text{1D})}=11\) with \(d^{(\text{c})}=\lambda/2\), and (A) a sensing array of \(N^{(\text{s},\text{1D})}=9\) with \(d^{(\text{s})}=\lambda/2\), or (B) a sparse sensing array of \(N^{(\text{s},\text{1D})}=3\) with \(d^{(\text{s})}=2\lambda\).}
    \label{fig:sumco_asy}
    % \todo[inline]{MH: figure is very nice, but the quintessence, i.e., that they result in the same sum co-array, is a bit lost in the caption and could be highlighted more clearly. 
    % \newline Since you don't mention the convolution at all, maybe it can make sense to omit the star operator?
    % \newline AF: added sum co array text in figure + correlation comment in caption}
    \vspace{-4mm}
\end{figure}

It has been proven that similar angular capabilities, hence \glspl{psf}, can be achieved with fewer elements.
However, to avoid grating lobes, communications hardware must operate in either transmit or receive mode with constant spacing close to $\lambda/2$~\cite{mandelli2022sampling}. 
As a result,  the transmit communications array is kept fixed with $\lambda/2$ spacing while the receive sensing part is optimized as a separate, almost co-located array. This approach differs from previous work that assumed a perfect full-duplex setup~\cite{Hu2023SparseFD}.

As an example, we recall the two different mono-static setups shown in Fig.~\ref{fig:sumco_asy}, both obtaining a sum co-array of \(N^{(\text{+},\text{1D})}=19\) with \(d^{(\text{s})}=\lambda/2\). The communication array is the same as previously introduced. However, case (B) with a sensing spacing of \(d^{(\text{s})}=2\lambda\) reduces the number of sensing array elements by nearly $90\%$ compared to the reference case (A) with \(d^{(\text{s})}=\lambda/2\).

Considering the beamforming operations of the sparse \(N^{(\text{s},\text{1D})}=3\) sensing array independently from the communications array, we can obtain the \gls{psf} shown in Fig.~\ref{fig:psf_joint_single} as the dotted line curve in purple color.
Here, we can see the grating lobes generated due to \(d^{(\text{s})}\gg\lambda/2\), which are to be avoided. 
However, in sensing what matters is the joint beamforming of the two arrays, whose \glspl{psf} is plot with solid lines.
We can also notice how the main lobe can be significantly narrowed as \(d^{(\text{s})}\) -- and thus also the aperture of the sum co-array -- is increased.

\begin{figure}
    \centering
    \input{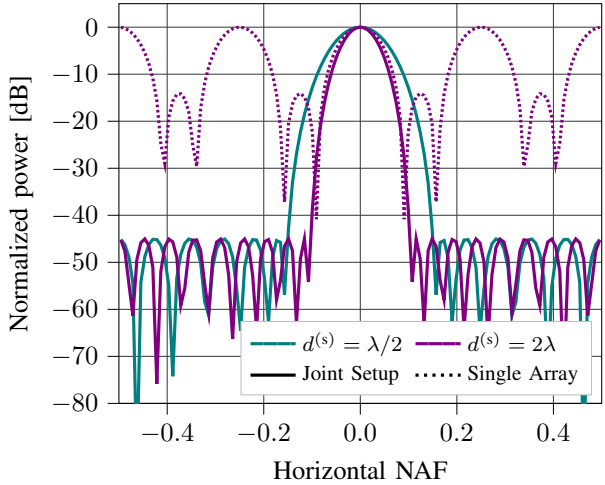}
    \caption{PSF comparison: Single sparse array with \(N^{(\text{s},\text{1D})}=3\) and \(d^{(\text{s})}=2\lambda\) vs. joint setup of fixed communications array with \(N^{(\text{c},\text{1D})}=11\) and \(d^{(\text{c})}=\lambda/2\) spacing and varying sensing array.}
    \label{fig:psf_joint_single}
    \vspace{-4mm}
\end{figure}

The example in the previous section highlights the potential of leveraging arrays with \(d^{(\text{s})}\gg\lambda/2\) in joint beamforming operations with conventional uniform arrays with \(d^{(\text{c})} = \lambda/2\). In Fig.~\ref{fig:Intensity_NAF_Sweep}, we extend our investigation to the \gls{2D} \gls{naf} domain by comparing the response of a scenario with two point scatterers, with three different \(d^{(\text{s})}\) values. We note a decreasing main lobe width when the antenna spacing \(d^{(\text{s})}\) increases. However, noise enhancement effects become noticeable when the spacing exceeds a certain threshold, specifically for (c) with \(d^{(\text{s})}=4\lambda\).
These noise effects are related to the lower redundancy in the sum co-array elements. Meaning less combinations of the Tx and Rx elements lead to the same sum co-array elements which translates to a higher dependency on each physical channel. It is therefore essential to identify a suitable trade-off between spacing and noise enhancement in practice.

\begin{figure*}[ht]
  \centering
  \hspace{-3mm}
  \subfloat
  		[\(d^{(\text{s})}=\lambda/2\). 
  		\label{fig:intensity_-10dB}]
  		{
		\pgfplotsset{scaled y ticks=false}

\def \globalscale {1}
\begin{tikzpicture}[yscale=\globalscale,xscale=\globalscale]

    \begin{axis}[
        axis on top,
        height=0.315\textwidth,
        width=0.315\textwidth,
        %scale only axis,
        %axis equal=true,
        grid=major,
        grid style={solid, black!15},
        enlargelimits=false,
        xmin=-0.5, xmax=0.5,
        xlabel near ticks,
        x label style={yshift=0.2em},
        ymin=-0.5, ymax=0.5,
        ytick={-0.5,-0.25,0,0.25,0.5},
        xtick={-0.5,-0.25,0,0.25,0.5},
        xticklabels={$-0.5$,$-0.25$,$0$,$0.25$,$0.5$},
        yticklabels={$-0.5$,$-0.25$,$0$,$0.25$,$0.5$},
        ylabel near ticks,
        xlabel={Horizontal \gls{naf}},
        ylabel={Vertical \gls{naf}},
        y label style={yshift=-2em},
        label style={font=\footnotesize},
        tick label style={font=\footnotesize},
        legend style={font=\footnotesize},
        legend pos = north east,
        legend style=
        	{fill=white, 
        	fill opacity=0.4, 
        	draw opacity=1, 
        	text opacity=1, 
        	nodes={scale=1, transform shape}, 
            /tikz/every even column/.append style={column sep=0.1cm}
        	},
        %colorbar horizontal,
        % colorbar,
        % point meta min=-60,
        % point meta max=0,
        colormap name=jet,
        % colorbar style=
        %     {ylabel={Power [dB]}, 
        %     at={(1.05, 0)},
        %     anchor = south,
        %     width = 0.3cm,
        %     ytick={-60,-40,-20, 0},
        %     ylabel style={yshift=0.8cm},
        %     every axis/.append style=
        %         {font=\footnotesize}
        %     }
        ]

      \addplot[forget plot] graphics[
      xmin=-0.5, 
      xmax=0.5, 
      ymin=-0.5, 
      ymax=0.5,
      includegraphics={
        trim=5 5 5 5,
        clip,}
        ] {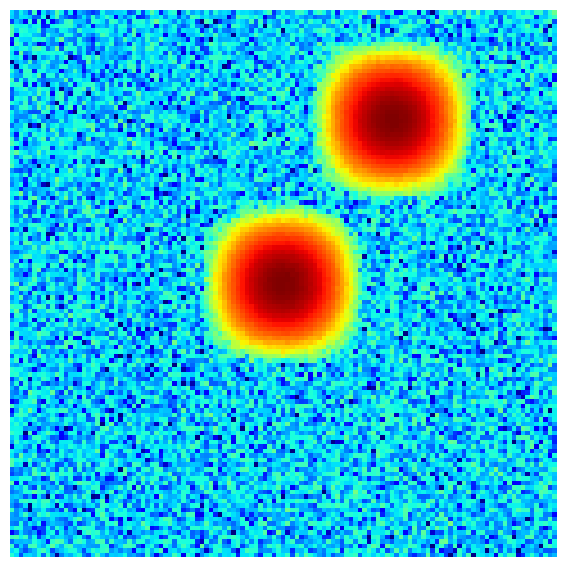};

    \end{axis}

\end{tikzpicture}
        \vspace{-1.5mm}
		} 
		\hspace{-6mm}
  \subfloat
  		[\(d^{(\text{s})}=2\lambda\).
  		\label{fig:intensity_0dB}]
  		{
		\pgfplotsset{scaled y ticks=false}

\def \globalscale {1}
\begin{tikzpicture}[yscale=\globalscale,xscale=\globalscale]

    \begin{axis}[
        axis on top,
        height=0.315\textwidth,
        width=0.315\textwidth,
        %scale only axis,
        %axis equal=true,
        grid=major,
        grid style={solid, black!15},
        enlargelimits=false,
        xmin=-0.5, xmax=0.5,
        xlabel near ticks,
        x label style={yshift=0.2em},
        ymin=-0.5, ymax=0.5,
        ytick={-0.5,-0.25,0,0.25,0.5},
        xtick={-0.5,-0.25,0,0.25,0.5},
        xticklabels={$-0.5$,$-0.25$,$0$,$0.25$,$0.5$},
        yticklabels={$-0.5$,$-0.25$,$0$,$0.25$,$0.5$},
        ylabel near ticks,
        xlabel={Horizontal \gls{naf}},
        ylabel={Vertical \gls{naf}},
        y label style={yshift=-2em},
        label style={font=\footnotesize},
        tick label style={font=\footnotesize},
        legend style={font=\footnotesize},
        legend pos = north east,
        legend style=
        	{fill=white, 
        	fill opacity=0.4, 
        	draw opacity=1, 
        	text opacity=1, 
        	nodes={scale=1, transform shape}, 
            /tikz/every even column/.append style={column sep=0.1cm}
        	},
        %colorbar horizontal,
        % colorbar,
        % point meta min=-60,
        % point meta max=0,
        colormap name=jet,
        % colorbar style=
        %     {ylabel={Power [dB]}, 
        %     at={(1.05, 0)},
        %     anchor = south,
        %     width = 0.3cm,
        %     ytick={-60,-40,-20, 0},
        %     ylabel style={yshift=0.8cm},
        %     every axis/.append style=
        %         {font=\footnotesize}
        %     }
        ]

      \addplot[forget plot] graphics[
      xmin=-0.5, 
      xmax=0.5, 
      ymin=-0.5, 
      ymax=0.5,
      includegraphics={
        trim=5 5 5 5,
        clip,}
        ] {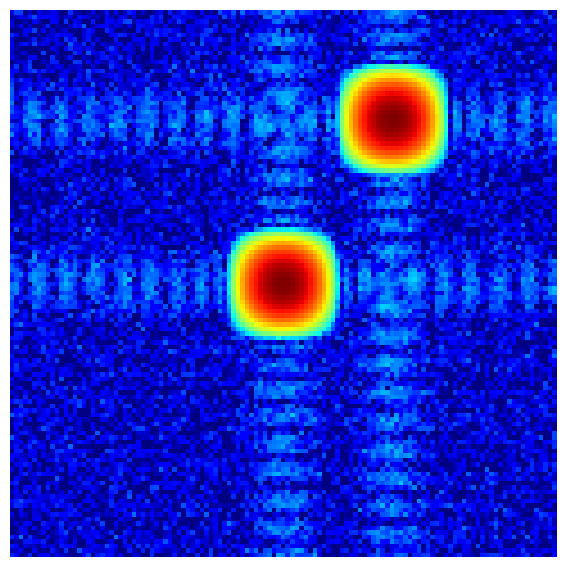};

    \end{axis}

\end{tikzpicture}
        \vspace{-1.5mm}
		} 
		\hspace{-6mm} %\\[5mm]
  \subfloat
  		[\(d^{(\text{s})}=4\lambda\).
  		\label{fig:intensity_10dB}]
  		{
		\pgfplotsset{scaled y ticks=false}

\def \globalscale {1}
\begin{tikzpicture}[yscale=\globalscale,xscale=\globalscale]

    \begin{axis}[
        axis on top,
        height=0.315\textwidth,
        width=0.315\textwidth,
        %scale only axis,
        %axis equal=true,
        grid=major,
        grid style={solid, black!15},
        enlargelimits=false,
        xmin=-0.5, xmax=0.5,
        xtick={-0.6,-0.4,-0.2,0,0.2,0.4,0.6},
        xlabel near ticks,
        x label style={yshift=0.2em},
        ymin=-0.5, ymax=0.5,
        ytick={-0.5,-0.25,0,0.25,0.5},
        xtick={-0.5,-0.25,0,0.25,0.5},
        xticklabels={$-0.5$,$-0.25$,$0$,$0.25$,$0.5$},
        yticklabels={$-0.5$,$-0.25$,$0$,$0.25$,$0.5$},
        ylabel near ticks,
        xlabel={Horizontal \gls{naf}},
        ylabel={Vertical \gls{naf}},
        y label style={yshift=-2em},
        label style={font=\footnotesize},
        tick label style={font=\footnotesize},
        legend style={font=\footnotesize},
        legend pos = north east,
        legend style=
        	{fill=white, 
        	fill opacity=0.4, 
        	draw opacity=1, 
        	text opacity=1, 
        	nodes={scale=1, transform shape}, 
            /tikz/every even column/.append style={column sep=0.1cm}
        	},
        %colorbar horizontal,
        colorbar,
        point meta min=-60,
        point meta max=0,
        colormap name=jet,
        colorbar style=
            {ylabel={Power [dB]}, 
            at={(1.10, 0)},
            anchor = south,
            width = 0.3cm,
            ytick={-60,-45,-15, 0},
            ylabel style={yshift=0.7cm},
            every axis/.append style=
                {font=\footnotesize}
            }
        ]

      \addplot[forget plot]graphics[
      xmin=-0.5, 
      xmax=0.5, 
      ymin=-0.5, 
      ymax=0.5,
      includegraphics={
        trim=5 5 5 5,
        clip,}
        ] {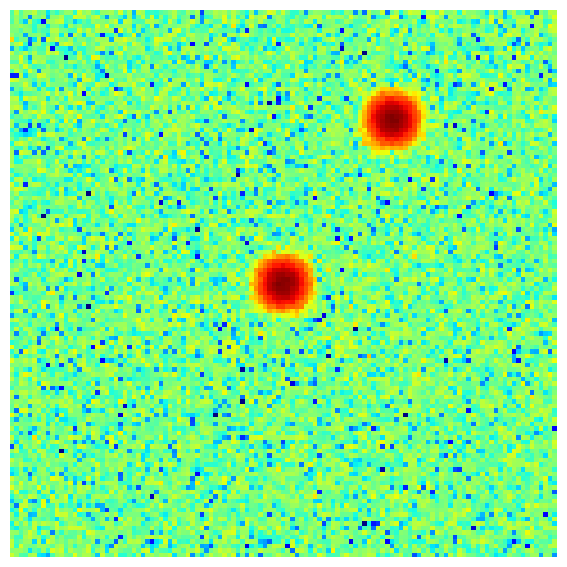};

    \end{axis}

\end{tikzpicture}
        \vspace{-1.5mm}
		} 
		%\quad
  \caption {Intensity plot over 2D \gls{naf} for $\sigma^2=-10$ dB observing 2 point scatterers for different joint arrays. The communications array is fixed with \(N^{(\text{c},\text{1D})}=11\) and \(d^{(\text{c})}=\lambda/2\), while the sensing array has fixed \(N^{(\text{s},\text{1D})}=3\) but varying \(d^{(\text{c})}\).}  
\label{fig:Intensity_NAF_Sweep}
\vspace{-4mm}
\end{figure*}

\section{Simulation Studies}

We perform a Monte Carlo simulation study to compare the angular capabilities of a mono-static setup as a function of the sensing array design. 
A $11\times 11$ \gls{ura} communication array, hence $N^\text{(c,\text{1D})}=11$, with elements of $d^\text{(c)}=\lambda/2$ spacing is considered, according to the number of elements and typical designs considered in 5G deployments~\cite{mandelli2023survey}
The sensing panel used as receiver varies in the number of antenna elements $N^{\text{(s,\text{1D})}}=\{3, 5, 7\}$, each size combined with three different element spacings $d^{\text{(s)}}=\{\lambda/2, \, 3\lambda/2, \, 2\lambda\}$.  More details about the simulation parameters are given in Table \ref{tab:sim_param}.

Each simulated scenario consists of two scatterers. The first scatterer is placed randomly on the \gls{2D} \gls{naf} plane, while a second is placed with a fixed \gls{2D} \gls{naf} distance $\Delta d^{(\text{P})}$, but at a random direction relative to the first. By varying $\Delta d^{(\text{P})}$, we analyze the ability of the setup to discriminate between multiple targets, i.e., its resolution.

To reconstruct the image of a scenario, we use the transposed representation of Eq. \eqref{eq:monostatic_imaging} in matrix operations. A single acquisition of the joint transmitter and receiver beamforming task is denoted by index $q$. For the $q$-th acquisition, acquired in the desired scan direction $\mathbf{u}$, this yields 
\begin{equation} \label{eq:monostatic_imaging_mat_operations}
y_q(\mathbf{u}) = \mathbf{w}_q^{(\text{s})\mathrm{T}}(\mathbf{u})\mathbf{A}^{(\text{s})} \mathbf{\Gamma} \mathbf{A}^{(\text{c})\mathrm{T}} \mathbf{w}_q^{(\text{c})}(\mathbf{u}) + \mathbf{w}_q^{(\text{s})\mathrm{T}}(\mathbf{u}) \mathbf{n}_q \; ,
\end{equation}
where the complex circular \gls{awgn} $\mathbf{n}_q$ has zero mean and as covariance matrix a scaled $2 \times 2$ identity matrix multiplied by $\sigma^2$. 
Further, the coefficients of the $K$ scatterers with constant amplitude and uniform random phase in the interval $[0, 2\pi)$ are represented by a diagonal matrix $\mathbf{\Gamma}\in \mathbb{C}^{K\times K}$. A \textit{pixel} of the complete image acquired in the direction $\mathbf{u}$ is obtained by the summation of all $Q$ joint beamforming acquisitions focused at $\mathbf{u}$
\begin{equation} \label{eq:monostatic_imaging_mat_operations_total}
y(\mathbf{u}) = \sum_{q=1}^{Q} y_q(\mathbf{u}) \; .
\end{equation}
Once the reconstructed image of the scenario has been acquired, targets are detected using \gls{cfar} thresholding with a desired probability of false alarm $P^\text{(FA)}=10^{-3}$.

Peaks above the \gls{cfar} threshold are detected and sorted by strength. One by one, their estimation is improved using interpolation techniques. With their estimated position, their effect is removed from the image using the techniques described in the multiple targets section (V-C) of~\cite{mandelli2022sampling}.

In our simulations, we measure performance in terms of target missed detection probability $P^\text{(MD)}$.
For a peak to be considered a detection, it must fall within an interval $\rho$ around the a true target position. The value of $\rho$ corresponds to the resolution of the considered mono-static setup. 
Note that we used a Chebyshev window with $\Omega=45$ dB sidelobe attenuation as the desired set of beamforming coefficients, resulting in a sidelobe amplitude of $r=\sfrac{1}{\left(10^{\Omega/20}\right)}$~\cite{Lynch1997Chebyshev}. Therefore, the half main-lobe width of the angular frequency response of the arrays in a single direction, is given by
\begin{equation} \label{eq:half-main-lobe-width}
\omega^{(+,\text{1D})}=\arccos{\left( \sfrac{1}{\mathrm{cosh}{\left(\frac{\mathrm{arccosh}{\left(\sfrac{1}{r}\right)}}{N^{(+,\text{1D})}-1}\right)}} \right)} \; .
\end{equation}
The resolution can be obtained by scaling the half main-lobe width with the limit of the respective \gls{naf} dimension, here:
\begin{equation} \label{eq:resolution}
\rho^{(+,\text{1D})}= 0.5 \, \omega^{(+,\text{1D})} \; .
\end{equation}

\renewcommand{\arraystretch}{1.30} %increase vertical space of table cells
\begin{table}
    \centering
    \caption{Simulation parameters and assumptions}
    \label{tab:sim_param}
     \begin{tabular}{|c|c|}
        \hline
        \textbf{Parameter} & \textbf{Value / Description} \\
        \Xhline{3\arrayrulewidth}
        Communications array & URA with $N^\text{(c,\text{1D})}=11$ and $d^\text{(c)}=\lambda/2$ \\
        % \hline
        % Communications array spacing $d^\text{(c)}$ & $\lambda/2$ \\
        \hline
        Number of targets & 2 \\
        \hline
        Target reflection coefficients & \(e^{j\Phi_s}\), where \(\Phi_s\backsim U\left[0,2\pi\right)\) \\
        \hline
        Distribution of coefficients  & Chebyshev $45$ dB attenuation \\
        % \hline
        % Detection method & Interpolated CFAR  \\
        % & Target cancellation as in~\cite{mandelli2022sampling} \\
        \hline
        Desired $P^\text{(FA)}$ of CFAR & \(10^{-3}\) \\
        \hline
        Iterations per data point & \(10^{4}\) \\
        \hline
    \end{tabular}

\vspace{-4mm}

\end{table}

For the probability of missed detection $P^\text{(MD)}$ curves in Fig.~\ref{fig:Perf_NAF_Sweep}, the sensing \gls{ura} is deployed with different numbers of elements and antenna spacing, represented by different colors and line styles, respectively. 
It is evident that the curves break down at a specific distance between the scatterers $\Delta d^{(\text{P})}$, indicating the resolution capabilities of the respective setup.

\begin{figure}
    \centering
    % \centerline{{\hypersetup{hidelinks}\ref{common_legend}}}
    % This file was created with tikzplotlib v0.10.1.
\begin{tikzpicture}

\def\scale{1.0}  % .72

\begin{semilogyaxis}[
    xlabel={$\Delta d^{(\text{P})}$},
    ylabel={$P^\text{(MD)}$},
    % ylabel shift = -14pt,
    ylabel style={font=\footnotesize,at={(axis description cs:.-0.02,.5)},rotate=0,anchor=south},
    ytick={0.00001, 0.00002, 0.00003, 0.00004, 0.00005, 0.00006, 0.00007, 0.00008, 0.00009, 0.0001, 0.0002, 0.0003, 0.0004, 0.0005, 0.0006, 0.0007, 0.0008, 0.0009, 0.001, 0.002, 0.003, 0.004, 0.005, 0.006, 0.007, 0.008, 0.009, 0.01, 0.02, 0.03, 0.04, 0.05, 0.06, 0.07, 0.08, 0.09, 0.1, 0.2, 0.3, 0.4, 0.5, 0.6, 0.7, 0.8, 0.9, 1, 2, 3, 4, 5, 6},
    yticklabels={$10^{-5}$, , , , , , , , ,$10^{-4}$, , , , , , , , ,$10^{-3}$, , , , , , , , ,$10^{-2}$, , , , , , , , , , , , , , , , , , $10^{0}$},
    xlabel style={font=\footnotesize},
    yticklabel style={font=\footnotesize,xshift=2pt},  % xshift=3pt
    xticklabel style={
        /pgf/number format/precision=4,
        /pgf/number format/fixed, 
        font=\footnotesize},
    ymin=1e-2,
    ymax=1e-0,
    xmin=0.02,
    xmax=0.18,
    scaled x ticks = false,
    minor tick num=1,
    yminorticks = true,
    enlargelimits = false,
    % legend pos = south west,
    grid = both,
    grid style={solid, black!15},
    % legend columns = 8,
    %transpose legend,
%    		legend style={at={(0.4,0.5)}, font=\scriptsize},
    legend cell align={left},
    legend columns = 4,
    legend style={
      fill opacity=1.0,
      draw opacity=1,
      text opacity=1,
      at={(0.999,0.999)},  % single col: (0.275,0.03) original: (0.91,0.86)
      anchor=north east,
      draw=lightgray,
      font=\scriptsize
    },
    % legend style={
    % cells={anchor=west},
    % % draw=none,
    % fill=none,
    % font=\footnotesize
    % },
%		xticklabels = {$1 \cdot 10^{-3}$, ,$2 \cdot 10^{-3}$,},
    % tick label style={/pgf/number format/fixed},
%        ticklabel style={font=\footnotesize},
    % legend cell align={left},
    every axis plot/.append style={very thick},
    mark repeat={2},
    log origin y=infty,
    set layers=Bowpark,
    % legend to name = common_legend,
    scale = \scale
]

% % legend / single column
% \addlegendimage{white, fill=white} % Dummy entry for legend title
% \addlegendentry{\hspace{-.5cm}\(N_1^{(\text{s})}\)}
% \addplot[very thick, color=violet, solid] coordinates {(0, 1)}; \addlegendentry{\(3\)}
% \addplot[very thick, color=teal, solid] coordinates {(0, 1)}; \addlegendentry{\(5\)}
% \addplot[very thick, color=cyan, solid] coordinates {(0, 1)}; \addlegendentry{\(7\)}
% \addlegendimage{white, fill=white} % Dummy entry for legend title
% \addlegendentry{\hspace{-.5cm}\(d^{(\text{s})}\)}
% \addplot[very thick, color=black, solid] coordinates {(0, 1)}; \addlegendentry{\( \lambda/2\)}
% \addplot[very thick, color=black, dashed] coordinates {(0, 1)}; \addlegendentry{\( 3\lambda/2\)}
% \addplot[very thick, color=black, dotted] coordinates {(0, 1)};\addlegendentry{\(2 \lambda\)}

% legend / single row
\addlegendimage{white, fill=white} % Dummy entry for legend title
\addlegendentry{\hspace{-.5cm}\(N^{(\text{s},\text{1D})}\)}
\addplot[very thick, color=violet, solid] coordinates {(0, 1)}; \addlegendentry{\(3\)\hspace{-.5cm}}
\addplot[very thick, color=teal, solid] coordinates {(0, 1)}; \addlegendentry{\(5\)}
\addplot[very thick, color=cyan, solid] coordinates {(0, 1)}; \addlegendentry{\(7\)}
\addlegendimage{white, fill=white} % Dummy entry for legend title
\addlegendentry{\hspace{-.5cm}\(d^{(\text{s})}\)}
\addplot[very thick, color=black, solid] coordinates {(0, 1)}; \addlegendentry{\( \lambda/2\)}
\addplot[very thick, color=black, dashed] coordinates {(0, 1)}; \addlegendentry{\( 3\lambda/2\)}
\addplot[very thick, color=black, dotted] coordinates {(0, 1)};\addlegendentry{\(2 \lambda\)}

\def\noise{"-10"}

\def\sensshape{"3"}
\def\sensspacing{"0.5"}
\addplot [very thick, violet, solid] plot table[x index=0, y index=1] {Data/Prob_MD/prob_md_\sensshape_by_\sensshape_d_\sensspacing_noise_\noise_dB.txt};
\def\sensspacing{"1.5"}
\addplot [very thick, violet, dashed] plot table[x index=0, y index=1] {Data/Prob_MD/prob_md_\sensshape_by_\sensshape_d_\sensspacing_noise_\noise_dB.txt};
\def\sensspacing{"2.0"}
\addplot [very thick, violet, dotted] plot table[x index=0, y index=1] {Data/Prob_MD/prob_md_\sensshape_by_\sensshape_d_\sensspacing_noise_\noise_dB.txt};

\def\sensshape{"5"}
\def\sensspacing{"0.5"}
\addplot [very thick, teal, solid] plot table[x index=0, y index=1] {Data/Prob_MD/prob_md_\sensshape_by_\sensshape_d_\sensspacing_noise_\noise_dB.txt};
\def\sensspacing{"1.5"}
\addplot [very thick, teal, dashed] plot table[x index=0, y index=1] {Data/Prob_MD/prob_md_\sensshape_by_\sensshape_d_\sensspacing_noise_\noise_dB.txt};
\def\sensspacing{"2.0"}
\addplot [very thick, teal, dotted] plot table[x index=0, y index=1] {Data/Prob_MD/prob_md_\sensshape_by_\sensshape_d_\sensspacing_noise_\noise_dB.txt};

\def\sensshape{"7"}
\def\sensspacing{"0.5"}
\addplot [very thick, cyan, solid] plot table[x index=0, y index=1] {Data/Prob_MD/prob_md_\sensshape_by_\sensshape_d_\sensspacing_noise_\noise_dB.txt};
\def\sensspacing{"1.5"}
\addplot [very thick, cyan, dashed] plot table[x index=0, y index=1] {Data/Prob_MD/prob_md_\sensshape_by_\sensshape_d_\sensspacing_noise_\noise_dB.txt};
\def\sensspacing{"2.0"}
\addplot [very thick, cyan, dotted] plot table[x index=0, y index=1] {Data/Prob_MD/prob_md_\sensshape_by_\sensshape_d_\sensspacing_noise_\noise_dB.txt};

\end{semilogyaxis}

\end{tikzpicture}
    \caption{Probability of missed detection at $\sigma^2=-10\text{dB}$ indicating the resolution capabilities of different mono-static setups. The communications array has fixed \(N^{\text{(c,\text{1D})}}=11\) with \(d^{\text{(c)}}=\lambda/2\) and the sensing array is varied according to legend.} 
    \label{fig:Perf_NAF_Sweep}
    \vspace{-4mm}
\end{figure}
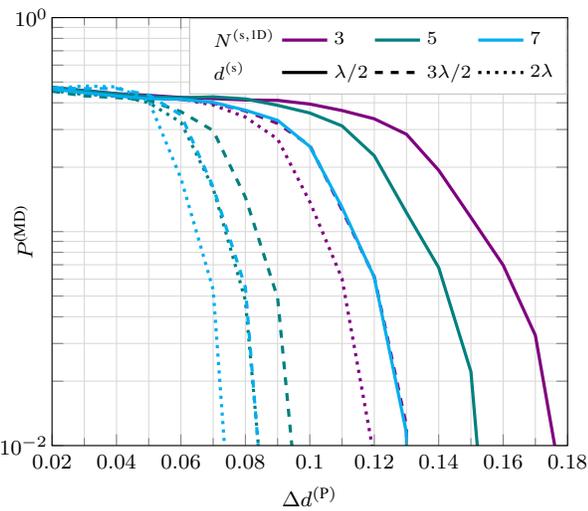

Performance improves as the number of elements increases, improving the resolution limit to a lower $\Delta d^{(\text{P})}$.
Comparing the performance of different antenna spacing, it is apparent that setups with greater $d^\text{(s)}$ outperform arrays with more elements but smaller spacing. 
For example, the dotted violet curve of $N^{(s,\text{1D})}=3$ and $d^\text{(s)}=2\lambda$ outperforms the solid green curve of $N^{(s,\text{1D})}=5$ and $d^\text{(s)}=\lambda/2$. 
These findings confirm the advantage of sensing arrays with fewer and sparse elements, requiring less computational complexity and ultimately resulting in cheaper solutions while still achieving the same resolution. Alternatively, within the same number of elements (identical color), an increase in spacing can significantly enhance the angular capabilities while maintaining similar costs.

% \balance
\section{Conclusion}

This work leverages the sparse array theory of radar to improve the design of mono-static 6G \gls{isac} setups.
Our proposal features a dissimilar array setup: one array with sparse element distribution and the other unchanged. We optimize the design of a half-duplex sensing array, that is coupled to a legacy half-duplex array designed for communications. 
With a simulation study, we have shown how the angular capabilities of the mono-static setups can be improved by a proper choice of the antenna spacing of the sensing array. 
The joint setup operates with fewer antenna elements in the sensing array and achieves the same angular capabilities as a full array, resulting in a better performance-to-cost trade-off.

As a next step, we plan an experimental evaluation using our \gls{isac} demonstrator in the ARENA2036~\cite{wild2023integrated}.

\balance
\appendices
% \section{First proof}


% \section{Second proof}
\section*{Acknowledgment}
This work was developed within the KOMSENS-6G project, partly funded by the German Ministry of Education and Research under grant 16KISK112K.
\else

\appendices
% \section{First proof}


% \section{Second proof}

\fi

\bibliographystyle{IEEEtran}
\bibliography{references.bib}

\begin{thebibliography}{10}
\providecommand{\url}[1]{#1}
\csname url@samestyle\endcsname
\providecommand{\newblock}{\relax}
\providecommand{\bibinfo}[2]{#2}
\providecommand{\BIBentrySTDinterwordspacing}{\spaceskip=0pt\relax}
\providecommand{\BIBentryALTinterwordstretchfactor}{4}
\providecommand{\BIBentryALTinterwordspacing}{\spaceskip=\fontdimen2\font plus
\BIBentryALTinterwordstretchfactor\fontdimen3\font minus \fontdimen4\font\relax}
\providecommand{\BIBforeignlanguage}[2]{{%
\expandafter\ifx\csname l@#1\endcsname\relax
\typeout{** WARNING: IEEEtran.bst: No hyphenation pattern has been}%
\typeout{** loaded for the language `#1'. Using the pattern for}%
\typeout{** the default language instead.}%
\else
\language=\csname l@#1\endcsname
\fi
#2}}
\providecommand{\BIBdecl}{\relax}
\BIBdecl

\bibitem{viswanathan2020communications}
H.~Viswanathan and P.~E. Mogensen, ``{Communications in the 6G era},'' \emph{IEEE Access}, vol.~8, pp. 57\,063--57\,074, Aug. 2020.

\bibitem{mandelli2023survey}
S.~Mandelli, M.~Henninger, M.~Bauhofer, and T.~Wild, ``Survey on integrated sensing and communication performance modeling and use cases feasibility,'' in \emph{2023 2nd International Conference on 6G Networking (6GNet)}, Oct. 2023, pp. 1--8.

\bibitem{liu2022survey}
A.~Liu \emph{et~al.}, ``A survey on fundamental limits of integrated sensing and communication,'' \emph{IEEE Communications Surveys \& Tutorials}, vol.~24, no.~2, pp. 994--1034, Feb. 2022.

\bibitem{wild2023integrated}
T.~Wild, A.~Grudnitsky, S.~Mandelli, M.~Henninger, J.~Guan, and F.~Schaich, ``{6G Integrated Sensing and Communication: From Vision to Realization},'' in \emph{2023 20th European Radar Conference (EuRAD)}, Sep. 2023, pp. 355--358.

\bibitem{hoctor1990unifying}
R.~T. Hoctor and S.~A. Kassam, ``The unifying role of the coarray in aperture synthesis for coherent and incoherent imaging,'' \emph{Proceedings of the IEEE}, vol.~78, no.~4, pp. 735--752, Apr. 1990.

\bibitem{Yu2016AltMinAlgo}
X.~Yu, J.-C. Shen, J.~Zhang, and K.~B. Letaief, ``Alternating {Minimization} {Algorithms} for {Hybrid} {Precoding} in {Millimeter} {Wave} {MIMO} {Systems},'' \emph{IEEE Journal of Selected Topics in Signal Processing}, vol.~10, no.~3, pp. 485--500, Apr. 2016.

\bibitem{wang2022beamforming_riemannian}
B.~Wang, Z.~Cheng, and Z.~He, ``Manifold optimization for hybrid beamforming in dual-function radar-communication system,'' \emph{Multidimensional Syst. Signal Process.}, vol.~34, no.~1, p. 1–24, Sep. 2022.

\bibitem{rajamaki2020hybrid}
R.~Rajam{\"a}ki, S.~P. Chepuri, and V.~Koivunen, ``Hybrid beamforming for active sensing using sparse arrays,'' \emph{IEEE Transactions on Signal Processing}, vol.~68, pp. 6402--6417, Oct. 2020.

\bibitem{Koc2022HybridBeamFD}
A.~Koc, ``{Hybrid} {Beamforming} {Techniques} in {Full}-{Duplex}/{Half}-{Duplex} {Massive} {MIMO} {Wireless} {Communications},'' Ph.D. dissertation, McGill University, 2022.

\bibitem{Hu2023SparseFD}
D.~Hu, X.~Wei, M.~Xie, and Y.~Tang, ``A {Sparse} {Shared} {Aperture} {Design} for {Simultaneous} {Transmit} and {Receive} {Arrays} with {Beam} {Constraints},'' \emph{Sensors}, vol.~23, no.~12, p. 5391, Jan. 2023.

\bibitem{mandelli2022sampling}
S.~Mandelli, M.~Henninger, and J.~Du, ``Sampling and reconstructing angular domains with uniform arrays,'' \emph{IEEE Transactions on Wireless Communications}, Nov. 2022.

\bibitem{rajamaki2019analog}
R.~Rajam{\"a}ki, S.~P. Chepuri, and V.~Koivunen, ``Analog beamforming for active imaging using sparse arrays,'' in \emph{2019 53rd Asilomar Conference on Signals, Systems, and Computers}, 2019, pp. 1202--1206.

\bibitem{Lynch1997Chebyshev}
P.~Lynch, ``The {Dolph}–{Chebyshev} {Window}: {A} {Simple} {Optimal} {Filter},'' \emph{Monthly Weather Review}, vol. 125, no.~4, pp. 655--660, Apr. 1997.

\end{thebibliography}

\end{document}